\documentclass[twocolumn]{jpsj3}
\usepackage{graphicx, color}

\usepackage{color}
\usepackage{textcomp}
\usepackage{bm}

\def\runauthor{Online-Journal Subcommittee}

\title{%
Local Lattice Distortion Caused by Short Range Charge Ordering in LiMn$_2$O$_4$
}
\author{%
Katsuaki \textsc{Kodama}$^1$, Naoki \textsc{Igawa}$^1$, Shin-ichi \textsc{Shamoto}$^1$, Kazutaka \textsc{Ikeda}$^2$,  
Hidetoshi \textsc{Oshita}$^2$, Naokatsu \textsc{Kaneko}$^2$, Toshiya \textsc{Otomo}$^2$, 
and Kentaro \textsc{Suzuya}$^3$ }

\inst{%
$^{1}$Quantum Beam Science Directorate, Japan Atomic Energy Agency, Tokai, Ibaraki 319-1195, Japan \\
$^{2}$Institute of Materials Structure Science, High Energy Accelerator Research Organization (KEK), 
Tsukuba, Ibaraki 305-0801, Japan\\
$^{3}$J-PARC Center, Japan Atomic Energy Agency, Tokai, Ibaraki 319-1195, Japan \\
}

\recdate{\today}

\abst{%
We have performed powder neutron diffraction on $^7$Li-enriched sample of LiMn$_2$O$_4$ at 300~K.   
The crystal structure determined by Rietveld analysis is a cubic spinel with space group of $Fd\bar{3}m$ in which all 
Mn atoms are crystallograghically equivalent, consistent with many preceding studies.  
However, the atomic pair distrubution function (PDF) of this compound can not be fitted by the cubic structure 
with space group of $Fd\bar{3}m$ satisfactorily, and it can be reproduced by the orthorhombic structure with $Fddd$.  
It corresponds with the structure of charge ordered phase below about 260~K, indicating a short range charge ordering.  
In the local structure determined by PDF analysis, two types of MnO$_6$ octahedra with long and short atomic distances 
between Mn and O atoms exist and their Mn-O distances are almost consistent with the distances in the charge ordered 
phase.  From these results, valence electrons are localized at Mn sites like a glass even in the cubic phase, 
resulting in the non-metallic electrical conductivity.
}

\kword{PDF analysis, neutron diffraction, local structure, short range charge ordering}

\begin{document}
\maketitle

\section{Introduction}
Charge ordering observed in strongly correlated electron system is one of typical examples of self-organization 
phenomena caused by many-body effect.  It can be regarded as a crystallization of valence electrons.  
Because the charge ordering is accompanied with a periodic lattice distortion corresponding with the periodicity of 
the arrangement of the localized electrons, it can be detected as an appearance of a superlattice reflection 
in diffraction data.\cite{schiffer, ohwada, chen}  
However, in several materials which have the same band-filling as those of the charge ordered mateirals and 
exhibit metal-insulator transitions, the superlattice reflection and/or the structural phase transtion can not 
be observed in their insulating phases.  It is considered that in such materials, although the electrons are localized 
by their Coulomb repulsion similar to charge ordered state, the arrangement of the localized electrons 
does not have a long range ordering, causing the lattice distortion without long range periodicity.  
It can be regarded as a glass state of valence electrons.  
In comparison with the charge ordered state, detailed study on such glass-like state of the charge has not been 
performed because only the local probe measurements such as NMR and M$\ddot{\textrm{o}}$ssbauer spectroscopy can 
be applied.\cite{hanasaki, kuzushita}  
In this paper, we focus on LiMn$_2$O$_4$ as a candidate of such materials in which the valence electron is 
freezed like a glass.  
\par
LiMn$_2$O$_4$ has been studied as a candidate of cathod materials in secondary lithium ion batteries, 
\cite{thackeray1, thackeray2, bruce, goodenough, guyomard} and as a frustrated magnetic materials.\cite{mukai, kamazawa}   
This compound has a cubic spinel structure with space group of $Fd\bar{3}m$ at room temperature.  
Mn atom is surrounded by six O atoms and they form MnO$_6$ octahedron.  
All Li, Mn and O atoms are crystallorgraghically equivalent, respectively.  
With decreasing temperature, the compound exhibits a structural phase transition at around 260K.
\cite{RC, tomeno, shimakawa, wills, oikawa, piszora}  
The low temperature phase has an orthorhombic structure with space group of $Fddd$ and "$3a \times 3a \times a$" 
super cell relative to the cubic phase, where $a$ is lattice parameter of the cubic phase.\cite{RC, piszora}  
Five kinds of inequivalent Mn sites are included in the unit cell.   
From the bond valence sum calculation, valences of three Mn and two Mn sites are estimated to be about +3 and +4, 
and then the orthorhombic phase is in the charge ordered state.\cite{RC}  
However, even in the cubic phase, the temperature dependence of the electrical resistivity is not metallic 
although the metallic conductivity can be expected from the simple band picture 
since the valence band which consists of 3$d$ orbits is partically filled because of the averaged valence of 
+3.5 of Mn ions.\cite{shimakawa, sugiyama, molenda1, molenda2}  
These results suggest that the valence electrons are localized at Mn sites like a glass due to their Coulomb repulsion, 
and the arrangement of Mn$^{3+}$ and Mn$^{4+}$ ions does not have a long range ordering in the cubic phase.  
\par
The lattice distortion without a long range periodicity (local lattice distortion) is an evidence of the glass-like 
freezing of the valence electrons.  In order to investigate the local structure around Mn ion in the cubic phase, 
extended x-ray-absorption fine-structure (EXAFS) measurements have been performed.\cite{exafs1, exafs2}  
However, the results of EXAFS measurements do not give a clear answer on the existence of the local lattice distortion 
; all Mn-O distances of MnO$_6$ octahedra have same bond length in ref. 23 suggesting no local lattice distortion, 
whereas two kinds of Mn-O distance are reported in ref. 22 suggesting the local lattice distortion induced 
by the existence of Mn$^{3+}$ and Mn$^{4+}$.  
The EXAFS measurement can probe only short atomic distance, for example, first or second nearest neighboring 
atomic distances.  It can also probe the local structure only around the selected atoms.  
Since it is difficult to detect the local lattice distortion only from the Mn-O distances 
because of the small difference between Mn$^{3+}$-O and Mn$^{4+}$-O distances (about 0.1 $\mathrm{\AA}$), 
the other atomic distances, for example, O-O distance should also be observed.  
Then in order to investigate the local lattice distortion in this compound, 
atomic pair distribution function (PDF) which can detect all atomic pair correlation should be used.  
\par   
In this paper, we report the results of the PDF analysis on neutron powder diffraction data on LiMn$_2$O$_4$ 
at room temperature where the average crystal structure is the cubic with the space group of $Fd\bar{3}m$.  
The obtained atomic pair distribution function is fitted for the orthorhombic structure with space group of $Fddd$ 
much better than the cubic structure, indicating the existence of the local lattice distortion.  
The locally distorted structure has MnO$_6$ octahedra with long and short Mn-O distances 
which almost correspond with the distances of Mn$^{3+}$-O and Mn$^{4+}$-O, respectively, 
suggesting that the valence electrons are localized at Mn sites with short range periodicity like a glass.   

\section{Experiments}
Powder sample of $^7$LiMn$_2$O$_4$ for present neutron measurements was prepared by following method.  
Here, we prepared $^7$Li-enriched sample in order to avoid the neutron absorption by natural abundance of $^6$Li.  
Starting materials are powders of $^7$LiOH$\cdot$H$_2$O and (CH3COO)$_2$Mn$\cdot$H$_2$O with stoichiometric composition 
and they were mixed in 2-propanol.  The mixture was dried at 400~\textcelsius~for 1 hour to decarbonate.  
After the calcinations, the powder was ground and then heated at 850~\textcelsius~for 10 hours.  
The sample of about 1.8~g was used in neutron diffraction measurements.  
\par
The powder neutron diffraction data for the Rietveld and PDF analyses were collected 
by using the neutron total scattering spectrometer NOVA installed in the Japan Proton Accelerator Research Complex 
(J-PARC).  The powder sample of about 1.7~g was set in vanadium-nickle alloy holder with a diameter of 0.6~cm.  
The data were collected at room temperature for about 8 hours.  

\section{Results and Discussions}
\subsection{Analysis on averaged crystal structure by using Rietveld analysis}
Neutron powder diffraction pattern of $^7$LiMn$_2$O$_4$ obtained at 90~\textdegree~ bank of NOVA is shown in Fig. 1 
by crosses.  In the plotted pattern, the contaminations of background intensities from the sample cell and 
the spectrometer are subtracted and the neutron absorption effect is calibrated.  
\begin{figure}[tbh]
\centering
\includegraphics[width=8.5cm]{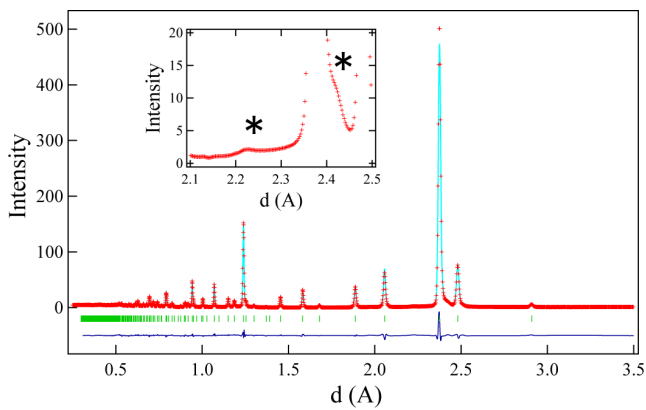} 
\caption{
(Color online) Observed (crosses) and calculated (solid line) neutron powder diffraction pattern of 
$^7$LiMn$_2$O$_4$.  Observed pattern is collected by using NOVA at room temperature.  
Vertical bars show the calculated position of Bragg reflections.  
The solid line at the bottom of the figure is the difference between observed and calculated intensities.  
Inset : foot of the main peak with $d \sim 2.4~\mathrm{\AA}$ are extended.  
}
\label{fig.1}
\end{figure}
Structural analysis on the neutron powder diffraction pattern is performed by using the program 
Z-Rietveld (ver.0.9.37.4).\cite{zcode1, zcode2}  
The reported space group of $Fd\bar{3}m$ is used.  
In the analysis, the occupation factors of Li and O atoms are also refined.
The obtained structural parameters of LiMn$_2$O$_4$ are shown in Table I.  
\begin{table}
\caption{
Atomic positions of LiMn$_2$O$_4$ determined by Rietveld analysis of neutron powder diffraction data 
at room temperature.  Space group of $Fd\bar{3}m$ (origin choice 2) is used in the analysis.  
Obtained lattice parameter is $a$=8.24333(2)~\AA. 
The $R$-factors, $R_\textrm{wp}$, $R_\textrm{p}$, $R_\textrm{e}$, $R_\textrm{B}$, and $R_\textrm{F}$ 
are 9.16~$\%$, 7.53~$\%$, 0.62~$\%$, 4.27~$\%$, 11.56~$\%$, respectively.
}
\label{t1}
\begin{center}
\begin{tabular}{lllllll}
\hline
Atom & Site & Occ. & $x$ & $y$ & $z$ & $B$~($\mathrm{\AA)}^{2}$) \\
\hline
Li & 8$a$ & 0.984(1) & 1/8 & 1/8 & 1/8 & 0.77(1) \\
Mn & 16$d$ & 1 & 1/2 & 1/2 & 1/2 & 0.61(1) \\
O & 32$e$ & 0.985(1) & 0.2634(1) & 0.2634 & 0.2634 & 0.99(1) \\
\hline
\end{tabular}
\end{center}
\end{table}
The errors of the parameters shown in the table are mathematical standard deviations obtained by the analysis.  
The diffraction pattern calculated by using refined parameters is shown in Fig. 1 by solid line.  
The calculated line reproduces the observed pattern.  
All Mn-O bonds are equivalent and the Mn-O distance is 1.957(1)~\AA, which is intermediate between what is expected 
for Mn$^{3+}$-O and Mn$^{4+}$-O distances.   
The deficiencies of Li and O sites are about 1.5~$\%$.  
If the analyis is performed for the occupation factors of Li and O fixed at 1.0, the fitting does not almost change. 
(For, example, $R_\textrm{wp}$ becomes 9.23 ~$\%$).  
Then the deficiencies of Li and O sites are almost negligible.  
Here, we can also neglect the possibility that excess Li atom occupies so-called $B$-site which is occupied 
by Mn atom because the present sample exhibits a structural phase transition from cubic to orthorhombic phase 
between 270~K and 240~K \cite{igawa} whereas the samples of Li$_{1+x}$Mn$_{2-x}$O$_4$ with $0 < x \lesssim 0.15$ 
retain the cubic symmetry in the whole temperature region.\cite{yamada, kamazawa}  
Although the analysis on the structure model with space group of $Fddd$ has also been carried out, 
meaningful improvement of fitting was not achieved.  
These results indicate that the present sample is almost stoichiometric LiMn$_2$O$_4$ and the averaged 
structure is the cubic with the space group of $Fd\bar{3}m$ at room temperature, as reported preceding studies.  
\par
In the inset figure, the foot of the main peak at $d\sim 2.4~\mathrm{\AA}$ is extended.  
Broad shoulder structures are observed at the foot of the main peak as shown by asterisks.  
The $d$ values of the broad shoulders at the larger and smaller $d$ sides almost correspond to the values of 2 10 0 
and 10 2 0, and 4 8 2 and 8 4 2 reflections in the $Fddd$ orthorhombic structure, respectively.  
These diffuse scatterings suggest the $Fddd$ orthorhombic lattice distortion with a short range correlation.

\subsection{Analyses on local structure by using PDF analysis}
Figures 2(a) and 2(b) show the structure function $S(Q)$ and the atomic pair distribution function $G(r)$ of 
$^7$LiMn$_2$O$_4$ at room temperature. 
The data are obtained from the neutron scattering intensity which is collected at back-scattering bank.  
$G(r)$ can be obtained by follwing Fourier transformation.
\begin{equation}
G(r)=\frac{2}{\pi} \int Q[S(Q)-1]sin(Qr)dQ.
\end{equation} 
In the present analysis, $S(Q)$ in the range of $1.21 \le Q \le 50$ $\mathrm{\AA}^{-1}$ is transformed into $G(r)$   
by using the program installed at NOVA.\cite{otomo}  
\begin{figure}[tbh]
\centering
\includegraphics[width=7.5 cm]{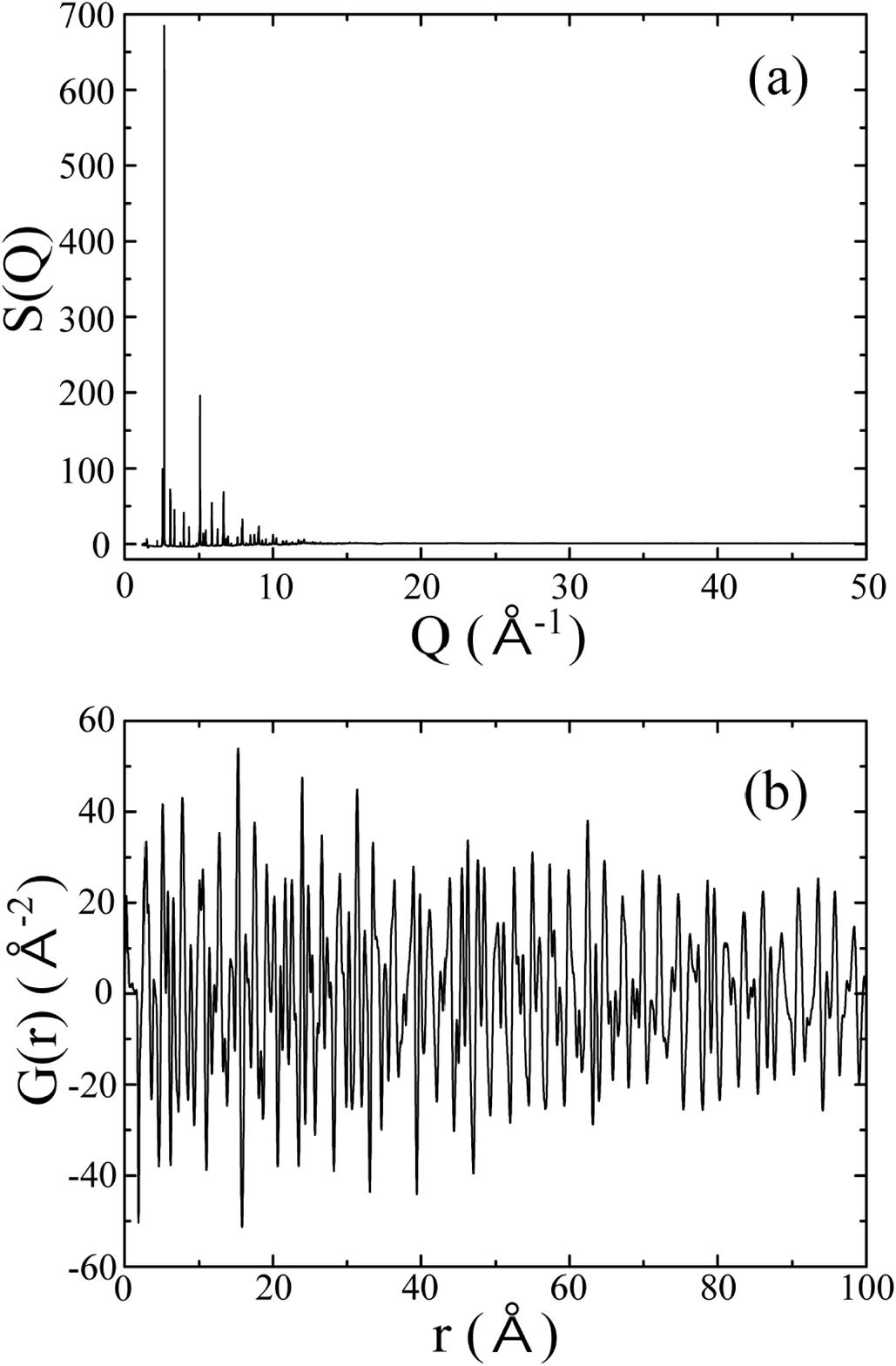} 
\caption{(a) The structure function $S(Q)$ obtained from the neutron diffraction data at room temperature.  
(b) The atomic pair distribution function (PDF) $G(r)$ obtained from $S(Q)$.}  
\label{fig.2}
\end{figure}
In Fig. 3(a), the fitting result by using the cubic structure with the space group of $Fd\bar{3}m$ 
which can reproduce the diffraction pattern in the previous subsection, is shown by solid line.  
In the analysis, occupation factors of all atoms are fixed at 1.0 because the deficiencies of Li and O atoms are 
only about 1.5~$\%$.  
The structural refinements on obtained $G(r)$ are performed by using the program PDFFIT.\cite{proffen2}  
The data in the region of $1.4 \le r \le 10~ \mathrm{\AA}$ are used for the fitting.  
\begin{figure}[tbh]
\centering
\includegraphics[width=8 cm]{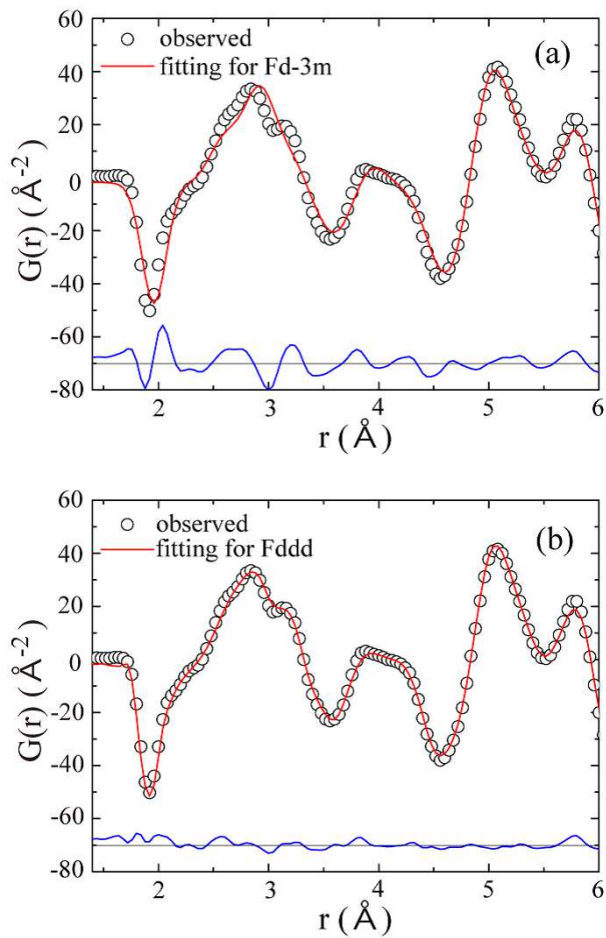} 
\caption{(Color online) Atomic pair distribution function (PDF) of $^7$LiMn$_2$O$_4$ obtained at room temperature 
(open circle).  Solid lines are the fittings for cubic (a) and orthorhombic (b) structures with space groups of 
$Fd\bar{3}m$ and $Fddd$, respectively.  Lines at lower position show the differences between observed data and 
fitting results.  Weighted $R$-factors obtained by fitting the data in the region of $1.4 \le r \le 10~ \mathrm{\AA}$ 
are 15.4~$\%$ and 6.94~$\%$ for cubic and orthorhombic structures, respcetively.}
\label{fig.3}
\end{figure}
The fitting line roughly reproduces the observed data.  However, the shapes of the first negative peak at about 
1.9~$\mathrm{\AA}$ and positive peaks around 2.8~$\mathrm{\AA}$ can not be reproduced by the line.  
The first negative peak almost consists of Mn-O atomic correlation in MnO$_6$ octahedron. Although Li-O correlation 
also contributes to the negative peak, the intensity is about 1/8 of the intensity of Mn-O correlation.  
The second positive peaks are superposition of Mn-Mn correlation between neighboring MnO$_6$ octahedra and O-O 
correlations which correspond to the correlation in the octahedron and the correlation between apical O atoms 
of the neighboring octahedra.  The intensity of Mn-Mn correlation peak is about 1/4 of the intensity of 
the O-O peaks.  In the cubic phase, the negative peak should be sharp and symmetric because the all Mn-O bonds 
are equivalent.  However, the negative peak has a shoulder structure at larger $r$ side, 
indicating the existence of the inequivalent Mn-O bonds.  Although, to fit the such asymmetric shape of the 
negative peak, the calculated negative peak becomes broad by adopting the large thermal factors of 
$B_\mathrm{Li}$=1.12, $B_\mathrm{Mn}$=0.947 and $B_\mathrm{O}$=1.11~$\mathrm{\AA}^2$ which are obtained by PDF 
analysis, the observed peak shape can not be reproduced by calculated $G(r)$.  
The Mn-Mn and O-O correlation peaks should be roughly symmetric three peak structure in the cubic phase 
because Mn-Mn distance is 2.92~$\mathrm{\AA}$ and O-O distances are 2.63, 2.92 and 3.21~$\mathrm{\AA}$ 
(the inensity ratio of O-O peaks is about 1:2:1).  
Because the large thermal factors broaden the calculated positive peaks of the Mn-Mn and O-O correlations, 
the calculated shape of the superposition of these positive peaks seems to be broad single peak around 
2.9~$\mathrm{\AA}$ and can not reproduce the observed complicated structure.   
The weighted $R$ factor, $R_\textrm{wp}$, obtained by fitting the data in the region of 
$1.4 \le r \le 10~ \mathrm{\AA}$ is 15.4~$\%$, indicating that the fitting is not satisfactory.  
\par
Then we use the orthorhombic structure with space group of $Fddd$ corresponding with the crystal structure 
in the charge ordered phase   
because the diffuse scattering which may be due to the $Fddd$ orthorhombic structure with short range correlation 
is observed in the diffraction pattern, as mention in the previous subsection.  
The fitting result by using the orthorhombic structure is shown in Fig. 3(b) 
by solid line.  
In this structure, five and nine inequivalent Mn and O atoms are contained in the unit cell, respectively.  
Because the atomic distances of 27 kinds of Mn-O bonds in MnO$_6$ octahedra distribute from about 1.82 to 
2.23~$\mathrm{\AA}$, shoulder structure of the first negative peak can be reproduced.  
At the same time, the calculated Mn-Mn and O-O correlation peaks can also reproduce the complicated peak structure 
around 2.8~$\mathrm{\AA}$. As a result, the fitting is improved and 
the $R_\textrm{wp}$ value is 6.94~$\%$ much smaller than the $R_\textrm{wp}$ obtained 
from the cubic structure.  In order to check the possibility of other local 
lattice distortions, fittings by using structure models with the maximal subgroups of $Fd\bar{3}m$ are performed.  
Figures 4(a), 4(b) and 4(c) show the fitting results by using the structure models with space groups of $F4\bar{3}m$, 
$I4_1/amd$ and $R\bar{3}m$, respectively.     
\begin{figure}[tbh]
\centering
\includegraphics[width=7 cm]{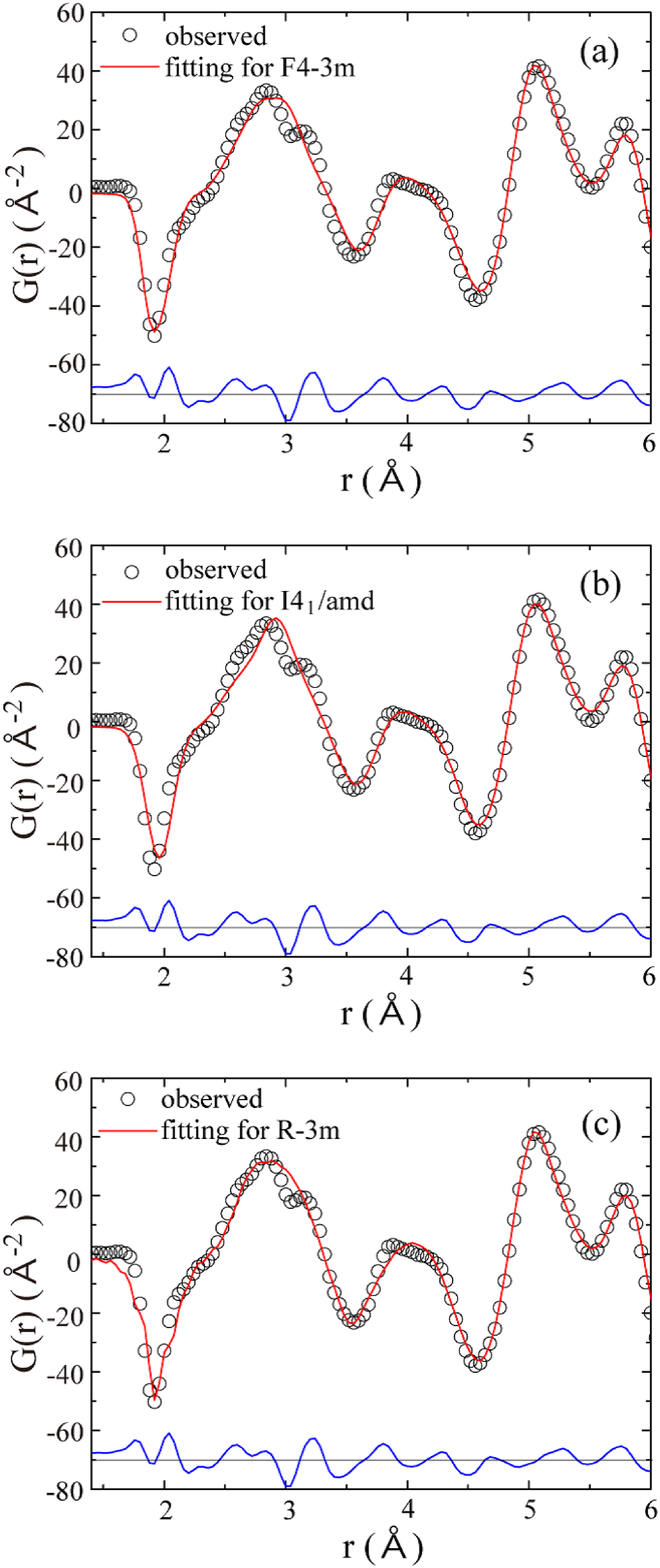} 
\caption{(Color online)Atomic pair distribution function (PDF) of $^7$LiMn$_2$O$_4$ obtained at room temperature 
(open circle).  Solid lines are the fittings for structures with space groups 
$F4\bar{3}m$ (a), $I4_1/amd$ (b) and $R\bar{3}m$ (c), respcetively.  Lines at lower position show the differences 
between observed data and fitting results.  Weighted $R$-factors obtained by fitting the data in the region of 
$1.4 \le r \le 10~ \mathrm{\AA}$ are 14.2~$\%$, 15.3~$\%$ and 15.2~$\%$ for $F4\bar{3}m$, $I4_1/amd$ and $R\bar{3}m$, 
respcetively.}
\label{fig.3}
\end{figure}
Here, space groups $F4_132$ and $Fd\bar{3}$ are not used for the fitting because the atomic sites 
of Li, Mn and O atoms in these space groups correspond with the sites in $Fd\bar{3}m$.  
The meaningful improvements of fitting are not achieved and $R$-factors do not 
becomes smaller significantly for the above three structural models despite the lower symmetries.  
The complicated shapes of the first negative and second positive peaks are not reproduced in these models although 
the structure models with $F4\bar{3}m$, $I4_1/amd$ and $R\bar{3}m$ contain two, two and three kinds of Mn-O bonds 
and four, five and seven kinds of O-O bonds, respectively.  
These results show that the cubic phase of LiMn$_2$O$_4$ has an orthorhombic local lattice distortion 
which corresponds with the structure in the charge ordered phase.  
The broad diffuse scattering observed in the diffraction pattern which is mentioned in the previous section, is consistent 
with the orthorhombic local lattice distortion detected by the present PDF analysis and 
indicates the short range ordering of this local lattice distortion.  
\par
In Table II, the structural parameters of the locally distorted structure of LiMn$_2$O$_4$ 
which are determined by the PDF analysis using the orthorhombic structure are shown.  
\begin{table}[tbh]
\caption{Atomic positions of LiMn$_2$O$_4$ determined by PDF analysis on neutron powder diffraction data 
at room temperature.  Space group of $Fddd$ (origin choice 2) 
is used in the analyis.  Occupation factors of all atoms are fixed at 1.0.  
Obtained lattice parameter is $a$=24.6388(3)~\AA, $a$=24.7978(3)~\AA, and $c$=8.21781(8)~\AA. 
The $R$-factor, $R_\textrm{wp}$, is 6.94~$\%$. }
\label{t2}
\begin{center}
\begin{tabular}{llllll}
\hline
Atom & Site  & $x$ & $y$ & $z$ & $B$~(\AA$^{2}$) \\
\hline
Li(1) & 8$a$ & 1/8 & 1/8 & 1/8 & 0.29(1) \\
Li(2) & 16$f$ & 1/8 & 0.7898(2) & 1/8 & 0.29 \\
Li(3) & 16$e$ & 0.7950(2) & 1/8 & 1/8 & 0.29 \\
Li(4) & 32$h$ & 0.2965(1) & 0.3016(1) & 0.1162(1) & 0.29 \\
Mn(1) & 16$d$  & 1/4 & 1/4 & 1/2 & 0.33(1) \\
Mn(2) & 32$h$  & 0.0833(1) & 0.0836(1) & 0.5083(1) & 0.33 \\
Mn(3) & 32$h$  & 0.0879(1) & 0.3298(1) & 0.2512(1) & 0.33 \\
Mn(4) & 32$h$  & 0.2517(1) & 0.1674(1) & 0.2505(1) & 0.33 \\
Mn(5) & 32$h$  & 0.1644(1) & 0.2458(1) & 0.2529(2) & 0.33 \\
O(1) & 32$h$  & 0.1744(1) & 0.1685(1) & 0.2585(2) & 0.56(1) \\
O(2) & 32$h$  & 0.0782(1) & 0.0049(1) & 0.4818(1) & 0.56 \\
O(3) & 32$h$  & 0.0785(1) & 0.3312(1) & 0.4756(1) & 0.56 \\
O(4) & 32$h$  & 0.2526(1) & 0.1721(1) & 0.4719(1) & 0.56 \\
O(5) & 32$h$  & 0.0035(1) & 0.0082(1) & 0.2487(2) & 0.56 \\
O(6) & 32$h$  & 0.2530(1) & 0.0897(1) & 0.2370(1) & 0.56 \\
O(7) & 32$h$  & 0.1627(1) & 0.3238(1) & 0.2370(2) & 0.56 \\
O(8) & 32$h$  & 0.0908(1) & 0.2467(1) & 0.2279(1) & 0.56 \\
O(9) & 32$h$  & 0.0843(1) & 0.1610(1) & 0.5150(1) & 0.56 \\
\hline
\end{tabular}
\end{center}
\end{table}
Here, the thermal factors of each atom are common in order to reduce the refined parameters.  In Table III, 
the atomic distances between Mn and O atoms averaged in MnO$_6$ octahedra determined from the parameters 
in Table II are shown.  
\begin{table}[tbh]
\caption{Averaged atomic distances between Mn and O atoms in MnO$_6$ octahedra and distortion parameters of MnO$_6$ 
octahedra, $\Delta$ obtained from the structural parameters in Table II.  Distrotion parameter of MnO$_6$ octahedron 
with averaged atomic distance $d$ is defined as $\Delta = 1/6 \Sigma_{n=1} ^6 [(d_n-d)/d]^2$.  
Values of $d$ and $\Delta$ determined by Rietveld analysis at 230~K \cite{RC} are also shown 
in the right side of the table.
}
\label{t3}
\begin{center}
\begin{tabular}{lllll}
\hline
& \multicolumn{2}{c}{
 PDF analysis at 300K} & \multicolumn{2}{c}{ Rietveld analysis at 230K\cite{RC} 
}  \\
\hline
& $d~\mathrm{\AA}$ & $\Delta$ ($\times 10^{-4}$) &  $d~\mathrm{\AA}$ &  $\Delta$ ($\times 10^{-4}$) \\
\hline
Mn(1)-O & 1.986(2) & 10.6 & 2.003(2)  & 20.6  \\
Mn(2)-O & 2.000(3) & 20.7 & 1.995(4)  &  19.4 \\
Mn(3)-O & 2.013(2) & 45.2 & 2.021(5) & 36.6 \\ 
Mn(4)-O & 1.895(3) & 5.9 & 1.903(4) &  4.6 \\
Mn(5)-O & 1.920(3) & 10.0 & 1.916(4) &  6.1 \\
\hline
\end{tabular}
\end{center}
\end{table}
The errors of the parameters shown in the tables are mathematical standard deviations obtained by the analysis.  
For comparison, the Mn-O distances in the orthorhombic phase determined by Rietveld analysis\cite{RC} are also 
shown in the right side of the table.  
The Mn(1)-O, Mn(2)-O and Mn(3)-O distances are about 2.00~$\mathrm{\AA}$, whereas the Mn(4)-O and Mn(5)-O distances 
are about 1.90~$\mathrm{\AA}$.  
The former and latter values are apparantly larger and smaller than the value of Mn-O distance in the averaged 
structure, 1.957(1)~$\mathrm{\AA}$, respectively.  Moreover, 
these values are very similar to the values obtained by Rietveld analysis on the charge ordered phase.\cite{RC}
From these results, the valences of Mn(1), Mn(2) and Mn(3) ions are about +3 whereas the valences of 
Mn(4) and Mn(5) ions are about +4.  Even in the cubic phase, valence electrons are localized at Mn sites 
similar to the case of the charge ordered phase with the orthorhombic structure.  
\par
Here, we emphasize that as mentioned in previous subsection, the averaged (periodic) structure at room temperature 
is the cubic structure with single Mn site.  Because the arrangement of valence electrons localized at Mn sites 
has only short range correlation, the lattice distortion caused by the localized electrons is not observed in the 
averaged structure determined by conventional structural analysis.  It can be regarded as a glass state of 
valence electrons whereas a charge ordered state can be regareded as a crystal state of the electrons.  
In the cases of amorphous alloys and metallic glasses, although a unique atomic arrangement with a short range ordering, 
for example, atomic distances and coordination numbers of the nearest neighbor atoms, is observed, 
the atomic arrangement does not have a long range periodicity.\cite{inoue}  
Similar situation may be achieved in the arrangement of the localized electrons in the cubic phase of LiMn$_2$O$_4$.  
Although the local arrangement of the electrons is unique and is consistent with the charge ordered state, 
the arrangement does not have a long range periodicity.  
The long range ordering of the arrangement of the electrons seems to be constricted by the geometrical frustration 
due to the atomic arrangement of Mn.  
In this state, the non-metallic electrical conductivity is compatible with the existence of only single Mn site with the 
valence of +3.5 in the averaged structure.  
Such glass-like state of valence electrons may be possible in other compounds with mixed valence states 
and non-metallic electrical conductivities.  
\par
In the orthorhombic phase, the distortion parameters, $\Delta$, of the octahedra including 
Mn(1), Mn(2) and Mn(3) whose valences are +3, are larger than the $\Delta$ of the octahedra including 
Mn(4) and Mn(5) with the valences of +4, as shown in the right side of Table III.  
Here, $\Delta$ is defined as $\Delta = 1/6 \Sigma_{n=1} ^6 [(d_n-d)/d]^2$.\cite{RC}  
This result indicates that
the charge ordering transition at about 260~K is accompanied with an orbital ordering at Mn$^{3+}$ sites due to 
Yahn-Teller effect.  
Distortion parameters of MnO$_6$ octahedra at room temperature obtained by the present PDF analysis are 
also shown in Table III of the left side.  
The relationship between the Mn valences and distortion parameters seems to be qualitatively consistent with 
the relationship in the chrage ordered phase; the distortion parameters of MnO$_6$ octahedra including Mn$^{3+}$ 
ions tend to be larger than those of MnO$_6$ octahedra including Mn$^{4+}$.  
It suggest the possibility 
that the short range orbital ordering of Mn$^{3+}$ ion also exist in the cubic phase of LiMn$_2$O$_4$.  
Such short range orbital ordering is also observed in LaMnO$_3$ 
at higher temperature than the orbital ordering temperature.\cite{qiu}  
However, as shown in the table, the distortion parameter of the octahedron of Mn(1) whose valence is +3, 
almost corresponds with the parameter of the octahedron of Mn(5) with the valence of +4, 
indicating that the relationship between the Mn valences and distortion parameters is incomplete at room temperature.  
The possibility of the short range orbital ordering of Mn$^{3+}$ ion should be revealed by, for example, 
the measurement just sbove the structural transition temperature.  
\par
The structural phase transition from the cubic to orthorhombic structures at around 260~K accompanied with the 
charge ordering and orbital ordering at Mn$^{3+}$ sites is first order.  
In first order phase transtion, the short range correlation of the low temperature phase is generally absent above 
the transition temperature.  However, as mentioned above, the short range cluster with the local lattice distortion 
corresponding with the low temperature phase is also observed in LaMnO$_3$.\cite{qiu}  
In LaMnO$_3$, the size of the cluster with the lattice distortion 
gradually develops with decreasing temperature, and discontinuosly vanishes accompanied with the first order 
transition.  In the present compound, the temperature dependence of the correlation length (periodicity) 
near the structural phase transition temperature should also be estimated by PDF analysis to confirm 
the first order transition.
\par
\section{Summary}
We have performed the neutron powder diffraction measurement on $^7$LiMn$_2$O$_4$ at room temperature. 
Although the averaged structure determined by the Rietveld analysis is the cubic spinel, 
the local structure determined by PDF analysis is the orthorhombic corresponding with 
the charge ordered phase.  The Mn-O distances of the locally distorted orthorhombic structure are almost 
consistent with the distances of Mn$^{3+}$-O and Mn$^{4+}$-O, indicating that the Mn$^{3+}$ and Mn$^{4+}$ sites 
are arranged with short range periodicity.  In the cubic phase of LiMn$_2$O$_4$, 
the valence electrons are localized like a glass at Mn sites, resulting in the non-metallic electrical conductivity.  

\section*{Acknowledgment}
The neutron scattering experiment was approved by the Neutron Scattering Program Advisory Committee of IMSS, KEK 
(Proposal No. 2009S06).
This work was supported by a Grant-in-Aid for Scientific Research (C) (24510129) from 
the Ministry of Education, Culture, Sports, Science and Technology, Japan.


\begin{thebibliography}{9}

\bibitem{schiffer}
P. Schiffer, A. P. Ramirez, W. Bao, and S.-W. Cheong : Phys. Rev. Lett. {\bf 75} (1995) 3336.  
\bibitem{ohwada}
K. Ohwada, Y. Fujii, N. Takesue, M. Isobe, Y. Ueda, H. Nakao, Y. Wakabayashi, Y. Murakami, K. Ito, Y. Amemiya, 
H. Fuhihisa, K. Aoki, T. Shobe, Y. Noda, and N. Ikeda : Phys. Rev. Lett. {\bf 87} (2001) 086402.  
\bibitem{chen}
C. H. Chen, S.-W. Cheong, and A. S. Cooper : Phys. Rev. Lett. {\bf 72} (2993) 2461.  
\bibitem{hanasaki}
N. Hanasaki, K. Masuda, K. Kodama, M. Matsuda, H. Tajima, J. Yamazaki, M. Takigawa, J. Yamaura, E. Ohmichi, 
T. Osada, T. Naito, and T. Inabe : J. Phys. Soc. Jpn. {\bf 75} (2006) 104713.
\bibitem{kuzushita}
K. Kuzushita, S. Morimoto, S. Nasu, and S. Nakamura : J. Phys. Soc. Jpn. {\bf 69} (2000) 2767.
\bibitem{thackeray1}
M. M. Thackeray, A. Dekock, M. H. Rossouw, D. Liles, R. Bittihn, and D. Hoge : J. Electrochem. Soc. {\bf 139} 
(1992) 363. 
\bibitem{thackeray2}
M. M. Thackeray : J. Electrochem. Soc. {\bf 142} (1995) 2558. 
\bibitem{bruce}
P. G. Bruce : Philos. Trans. R. Soc. London Ser. A {\bf 354} (1996) 1577.
\bibitem{goodenough}
J. B. Goodenough : Solid State Ionics {\bf 69} (1994) 184. 
\bibitem{guyomard}
D. Guyomard and J. M. Tarascon : Solid State Ionics {\bf 69} (1994) 222. 
\bibitem{mukai}
K. Mukai, J. Sugiyama, K. Kamazawa, Y. Ikedo, D. Andreica, and A. Amato : J. Solid State Chem. {\bf 184} (2011) 1096.
\bibitem{kamazawa}
K. Kamazawa, H. Nozaki, M. Harada, K. Mukai, Y. Ikedo, K. Iida, T. J. Sato, Y. Qiu, M. Tyagi, and J. Sugiyama : 
Phys. Rev. B {\bf 83} (2011) 094401.  
\bibitem{RC}
Rodriguez-Caravajal, G. Rousse, C. Masquelier, and M. Hervieu : Rhys. Rev. Lett. {\bf 81} (1998) 4660.  
\bibitem{tomeno}
I. Tomeno, Y. Kasuya, and Y. Tsunoda : Phys. Rev. B {\bf 64} (2001) 094422.
\bibitem{shimakawa}
Y. Shimakawa, T. Numata, and J. Tabuchi : J. Solid State Chem. {\bf 131} (1997) 138.
\bibitem{wills}
A. S. Wills, N. P. Raju, and J. E. Greedan : Chem. Mater. {\bf 11} (1999) 1510.
\bibitem{oikawa}
K. Oikawa, T. Kamiyama, F. Izumi, B. C. Chakoumakos, H. Ikuta, M. Wakihara, J. Li, and Y. Matsui : 
Solid State Ionics {\bf 109} (1998) 35. 
\bibitem{piszora}
P. Piszora : J. Alloy Compounds {\bf 382} (2004) 112.
\bibitem{sugiyama}
J. Sugiyama, T. Atsumi, A. Koiwai, T. Sasaki, T. Hioki, S. Noda, and N. Kamegashira : 
J. Phys. : Condens. Matter {\bf 9} (1997) 1729.  
\bibitem{molenda1}
J. Molenda, K. Swierczek, W. Kucza, J. Marzec, and A. Stoklosa : Solid State Ionics {\bf 123} (1999) 155. 
\bibitem{molenda2}
J. Molenda, K. Swierczek, M. Molenda, and J. Marzec : Solid State Ionics {\bf 135} (2000) 53. 
\bibitem{exafs1}
Y. Shiraishi, I. Nakai, T. Tsubota, T. Himeda, and F. Nishikawa : J. Solid State Chem. {\bf 133} (1997) 587.
\bibitem{exafs2}
A. Paolone, C. Cantellano, R. Cantelli, G. Rousse, and C. Masquelier : Phys. Rev. B {\bf 68} (2003) 014108.
\bibitem{zcode1}
R. Oishi, M. Yonemura, Y. Nishimaki, S. Torii, A. Hoshikawa, T. Ishigaki, T. Morishima, K. Mori, and T.Kamiyama : 
Nucl. Instrum. Method Phys. Res. Sect. A {\bf 94} (2009) 600. 
\bibitem{zcode2}
R. Oishi-Tomiyasu, M. Yonemura, T. Morishima, A. Hoshikawa, S. Torii, T. Ishigaki, and T. Kamiyama : 
J. Appl. Cryst. {\bf 45} (2012) 299.
\bibitem{igawa}
N. Igawa, K. Kodama, T. Taguchi, and S. Shamoto : Abstr. Meet. The Ceramics Society of Japan (24th Fall Meet., 2011), 
p. 443, 2P004 [in Japanese].
\bibitem{yamada}
A. Yamada : J. Solid State Chem. {\bf 122} (1996) 160.
\bibitem{otomo}
T. Otomo, K. Ikeda, H. Oshita, K. Suzuya, K. Itoh, T. Fukunaga, and K. Maruyama : private communication
\bibitem{proffen2} 
Th. Proffen and S. J. L. Billinge : J. Appl. Crystallogr. {\bf 32} (1999) 572.
\bibitem{inoue}
For example, A. Inoue : Acta mater. {\bf 48} (2000) 279.
\bibitem{qiu} 
X. Qiu, Th. Proffen, J. F. Mitchell, and S. J. L. Billinge : Phys. Rev. Lett. {\bf 94} (2005) 177203.  

\end{thebibliography}
\end{document}